\documentclass[aps,prb,twocolumn,superscriptaddress,amsmath,amssymb]{revtex4}
\usepackage{graphicx}
\usepackage{dcolumn}
\usepackage{bm}
\usepackage{color}
\usepackage{amsmath}
\usepackage{bbold}
\usepackage{graphicx, color}
\usepackage{amsfonts,amssymb,amsmath}
\usepackage{siunitx}
\usepackage{bm}
\usepackage{bbold}
\usepackage{dsfont}

\renewcommand{\vec}[1]{{\boldsymbol #1}}

\usepackage{letltxmacro}
\LetLtxMacro{\originaleqref}{\eqref}

\usepackage{etoolbox}
\pretocmd{\eqref}{Eq.~}{}{}
\usepackage{siunitx}

\makeatletter
\let\saved@includegraphics\includegraphics
\AtBeginDocument{\let\includegraphics\saved@includegraphics}
\renewenvironment*{figure}{\@float{figure}}{\end@float}
\makeatother

\begin{document}

\title{Quantum confinement of the Dirac surface states in topological-insulator nanowires}

\affiliation{Physics Institute II, University of Cologne, Z{\"u}lpicher Str. 77, 50937 K{\"o}ln, Germany}
\affiliation{Institute for Theoretical Physics, University of Cologne, Z{\"u}lpicher Str. 77, 50937 K{\"o}ln, Germany}
\affiliation{Institute of Physical Chemistry, University of Cologne, Luxemburger Str. 166, 50939 K\"oln, Germany}

\author{Felix M{\"u}nning}
\affiliation{Physics Institute II, University of Cologne, Z{\"u}lpicher Str. 77, 50937 K{\"o}ln, Germany}
\affiliation{These authors contributed equally}
\author{Oliver Breunig}
\affiliation{Physics Institute II, University of Cologne, Z{\"u}lpicher Str. 77, 50937 K{\"o}ln, Germany}
\affiliation{These authors contributed equally}
\author{Henry F. Legg}
\affiliation{Institute for Theoretical Physics, University of Cologne, Z{\"u}lpicher Str. 77, 50937 K{\"o}ln, Germany}
\affiliation{These authors contributed equally}
\author{Stefan Roitsch}
\affiliation{Institute of Physical Chemistry, University of Cologne, Luxemburger Str. 166, 50939 K\"oln, Germany}
\author{Dingxun Fan}
\affiliation{Physics Institute II, University of Cologne, Z{\"u}lpicher Str. 77, 50937 K{\"o}ln, Germany}
\author{Matthias R{\"o}{\ss}ler}
\affiliation{Physics Institute II, University of Cologne, Z{\"u}lpicher Str. 77, 50937 K{\"o}ln, Germany}
\author{Achim Rosch}
\affiliation{Institute for Theoretical Physics, University of Cologne, Z{\"u}lpicher Str. 77, 50937 K{\"o}ln, Germany}
\author{Yoichi Ando}
\affiliation{Physics Institute II, University of Cologne, Z{\"u}lpicher Str. 77, 50937 K{\"o}ln, Germany}

\maketitle 
%
%

{\bf The non-trivial topology of the three-dimensional (3D) topological insulator (TI) dictates the appearance of gapless Dirac surface states. Intriguingly, when a 3D TI is made into a nanowire, a gap opens at the Dirac point due to the quantum confinement, leading to a peculiar Dirac sub-band structure\cite{Zhang2010, Bardarson2010}. This gap is useful for, e.g., future Majorana qubits based on TIs\cite{Manousakis2017}. Furthermore, these Dirac sub-bands can be manipulated by a magnetic flux and are an ideal platform for generating stable Majorana zero modes (MZMs)\cite{Cook2011, Alicea2012}, which play a key role in topological quantum computing\cite{Nayak2008}. However, direct evidence for the Dirac sub-bands in TI nanowires has not been reported so far. Here we show that by growing very thin ($\sim$40-nm diameter) nanowires of the bulk-insulating topological insulator (Bi$_{1-x}$Sb$_x$)$_2$Te$_3$ and by tuning its chemical potential across the Dirac point with gating, one can unambiguously identify the Dirac sub-band structure. Specifically, the resistance measured on gate-tunable four-terminal devices was found to present non-equidistant peaks as a function of the gate voltage, which we theoretically show to be the unique signature of the quantum-confined Dirac surface states. These TI nanowires open the way to address the topological mesoscopic physics, and eventually the Majorana physics when proximitised by an $s$-wave superconductor.}

In TI nanowires\cite{Zhang2010, Bardarson2010, Cook2011}, the quantum confinement of the electron motion along the circumferential direction is described by the angular-momentum quantum number $\ell$.
In zero magnetic field, this quantization leads to the gap opening at the Dirac point, and the sub-bands become doubly-degenerate (see Fig. 1a). When a magnetic flux $\Phi$ threads along the wire, the energy spectrum is modified in an nontrivial way as described by the following formula (under the simplified assumption of a circular wire):   
\begin{equation}
 E_\ell(k)=\pm\hbar v_\mathrm{F} \sqrt{k^2+\left(\frac{\ell-\eta}{R_{\rm w}}\right)^2}, \quad \eta \equiv \Phi/\Phi_0. \label{eqn:Elk}
\end{equation}
Here, $v_\mathrm{F}$ is the Fermi velocity, $R_{\rm w}$ is the wire radius, and $\Phi_0=hc/e$ is the flux quantum; note that $\ell$ takes half-integer values $\pm \frac{1}{2},\pm \frac{3}{2},\dots$ due to a Berry phase arising from the spin-momentum locking of the TI surface states\cite{Zhang2010}.
Interestingly, a spin-non-degenerate gapless spectrum is restored when $\Phi$ is a half-integer multiple of $\Phi_0$; the spin-momentum locking in this gapless subband leads to the appearance of MZMs when the wire is proximitized by an $s$-wave superconductor\cite{Cook2011}. The tunability of the spin-momentum locking with $\Phi$ makes the sub-bands described by Eq. (1) a particularly interesting platform for topological mesoscopic physics. 

In experiments, to elucidate the peculiar quantization effects, the TI nanowire should be bulk-insulating and as narrow as possible, preferably less than $\sim$100 nm. Past efforts for TI nanowires\cite{Peng2009, Xiu2011, Tian2013, Hamdou2013, Safdar2013, Hong2014, Baessler2015, Cho2015, Arango2016, Jauregui2016, Bhatt2017, Ziegler2018} have only been able to indirectly probe the quantized Dirac sub-bands, although bulk-insulating TI nanowires have been occasionally reported\cite{Hong2012, Wang2013, Cho2015, Jauregui2015, Dufouleur2017, Kunakova2018, Ziegler2018}. In this work, we employed the vapour-liquid-solid (VLS) method using Au nanoparticles as catalysts\cite{Peng2009} and applied the concept of compensation, which has been useful for achieving bulk-insulation in bulk crystals\cite{Ren2011, Ando2013}.
Specifically, we tuned the Bi/Sb ratio of $\mathrm{(Bi_{1-x}Sb_x)_2Te_3}$ nanowires to a value that yields the most insulating properties. Fabrication of gate-tunable four-terminal devices allows us to bring the chemical potential across the Dirac point, upon which we discovered unusual oscillatory behaviour in the resistance near the Dirac point in very thin wires. This feature turns out to be the signature of the quantized Dirac sub-bands in TI nanowires as our theoretical calculations show.

During the VLS growth, the catalysts form a constantly over-saturated liquid alloy with the absorbed source materials, which then precipitate and form a crystal underneath. Using nominally-20-nm-diameter Au nanoparticles as catalysts, we obtain nanowires with a constant diameter between 20 to 100 nm, with a length of up to several $\mu$m (Fig. 1b). By using transmission electron microscopy (TEM) and energy-dispersive X-ray (EDX) analysis (Fig.~1c), we identify the Au catalyst at the tip of most of the analysed nanowires. The wires are found to be surrounded by a $\sim$4-nm-thick amorphous oxide shell.
The selected-area diffraction patterns (SAED, Fig. 1c inset) indicate a high crystalline quality. We found hexagonal symmetry for a direction perpendicular the nanowire axis, which allows us to identify the growth direction to be $\langle 11\bar{2}0 \rangle$-type. The compositional analysis using EDX along the wire shows a constant stoichiometry (Bi$_{0.68}$Sb$_{0.32}$)$_2$Te$_3$ in the nanowire core and no incorporation of Au was detected. (See Supplementary Information.)

\begin{figure*}[t]
\centering
\includegraphics[width=\textwidth]{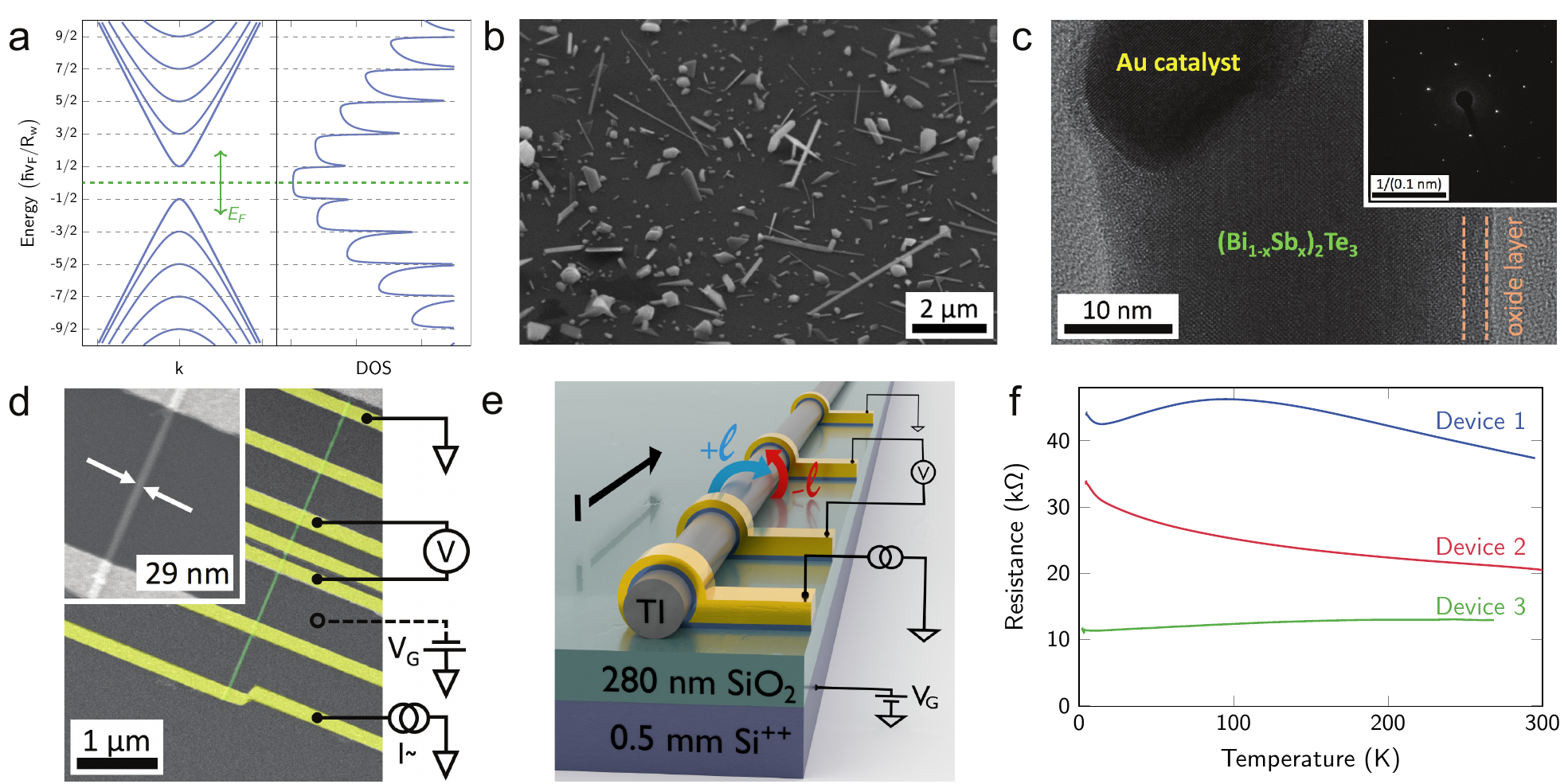}
\caption{ \textbf{Topological-insulator nanowire and its device.} \textbf{a}, The sub-band structure of quantum-confined TI surface states described by Eq.~\eqref{eqn:Elk} (left) and corresponding density of states (right). \textbf{b}, SEM image of nanowires (and other nanostructures) grown on a substrate. \textbf{c}, TEM image revealing the single-crystalline nanowire body, the remainder of the \SI{20}{nm} Au-nanoparticle growth catalyst, and a 4-nm thick oxide layer at the nanowire surface; inset shows the SAED pattern taken in the $c$-axis direction. TEM image and SAED pattern were taken from different nanowires and directions.
\textbf{d}, SEM micrograph of device 3 with the schematics of electrical wiring; inset show a magnified view of the nanowire. \textbf{e}, Schematic 3D image of the device construction. \textbf{f}, $R(T)$ curves of the three devices. The distance between the centers of the voltage contacts used for taking these data were 1.2, 0.5, and 0.6 $\mu$m, and the diameter of the nanowires were 43, 40, and 29 nm for devices 1, 2, and 3, respectively.}
\label{fig:1}
\end{figure*}

In the following, we report three representative devices 1 -- 3. The scanning electron microscope (SEM) picture of device 3 is shown in Fig. 1d, with its schematic depicted in Fig. 1e. The resistance $R$ vs. temperature $T$ curves shown in Fig. 1f present both insulating and metallic behaviour; nevertheless, all three samples were bulk-insulating, which can be seen in their gate-voltage $V_{\rm G}$ dependences of $R$ (Figs. 2a--2c) showing a clear maximum, indicating that the Dirac point is crossed. 
The difference in the $R(T)$ dependence is most likely explained by a slightly different electron density $n$
of the three samples in the absence of gating ($n \approx -0.2, +0.4, +0.5$~nm$^{-1}$ relative to the Dirac point, according to our analysis described later).
In the $R(V_{\rm G})$ traces, we found a hierarchy of fluctuation features. These fluctuations are not the universal conductance fluctuations (UCF), as they are much too regular. We observe semi-oscillatory features in the $V_{\rm G}$ dependence with the amplitude $A_\mathrm{I}\approx \SI{5}{k\Omega}$ (type I), which we will show to be the signature of sub-band crossings (for our methodology to identify type-I peaks, see Supplementary Information).
The other type of fluctuations has the amplitude $A_\mathrm{II}\approx$ 0.5 k$\Omega$ (type II) and was changing with time (Fig. S2 of the Supplementary Information). We speculate that they arise from time-dependent conductance fluctuations due to charge traps or mobile scattering centers, similar to those observed in metallic nanowires of similar mesoscopic size\cite{Beutler1987}, but they may also be affected by the presence of electron-hole puddles\cite{Borgwardt2016,Breunig2017,Boemerich2017}, see Supplementary Information.

\begin{figure*}[t]
\centering
\includegraphics[width=\textwidth]{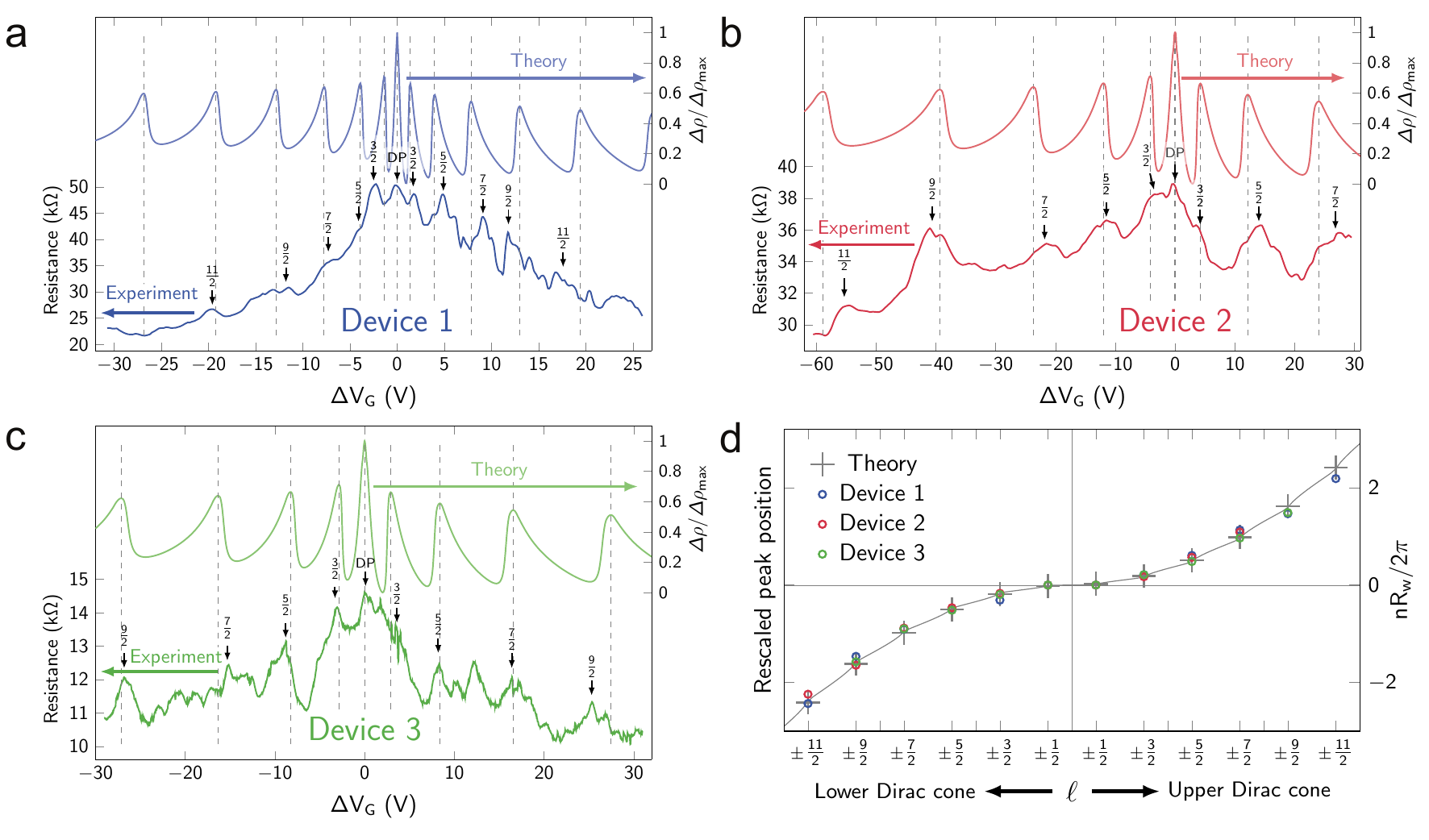}
\caption{\textbf{Signature of sub-band crossings}.\ 
In panels \textbf{a-c}, lower curves show the $V_{\rm G}$ dependence of $R$ observed in devices 1--3 at 2 K; $\Delta V_{\rm G}=0$ corresponds to the Dirac point of the TI surface state, which was achieved with $V_{\rm G}$ of $-6.0$, $30.5$, and $19$\,V in devices 1, 2, and 3, respectively. 
Large oscillations of type I (see main text) arise from the sub-band structure and the corresponding maxima are labeled by arrows. Oscillations of type II are much smaller and differ for up and down sweeps.
Upper curves in panels \textbf{a-c} are the theoretically-calculated resistivity with a small density of impurities, assuming that electron density $n$ is proportional to $V_{\rm G}$; we used $R_{\rm w}$ of 21.5, 20, and 14.5 nm, and $C_{\rm G}$ of $5.8$, $2.0$, and $4.1$\,pF$/$m for devices 1, 2, and 3, respectively. Pronounced maxima arise at sub-band crossings (dashed lines). 
\textbf{d}, Rescaled position of the resistivity maxima of the three devices ($\Delta V_{\rm G}/V_{0}$, see main text) as function of the quantum number $\ell$ compared to the theoretically calculated $n$ (in units of $2 \pi/R_{\rm w}$) at sub-band crossings.
}
\label{fig:2}
\end{figure*}

We now discuss the main observation of this work, that is, the reproducible semi-oscillatory feature in the $R(V_{\rm G})$ curves. Due to the 1D nature of the energy bands in the nanowire, the density of states (DOS) diverges as $1/\sqrt{E}$ at each of the sub-band's edges as shown in Fig. 1a. This causes a sub-band crossing to have two contrasting effects on $R$: (i) The opening of a new conductance channel can {\em decrease} the resistivity as more charge can be transported. It can, however, also (ii) {\em increase} the resistivity by opening a new channel where electrons from other bands can scatter into.
Thus, we have performed a straightforward theoretical calculation using an idealized model based on the surface state of a circular TI nanowire. The effects of local impurities is taken into account using the T-matrix formalism and we find that the experimental data is best described by weak impurities (see Supplementary Information). In Fig. 3, we schematically show how different sub-bands contribute to the conductivity: When a new channel is added (Fig. 3b), all other channels scatter efficiently into the new channel and, as a result, the conductivity contribution of each channel drops. This is by far the dominant effect and leads to pronounced peaks in $R$ even when several channels are present. The diverging density of states of the newly added channel (Fig.~1a) is the main reason why this effect is so large, but it is further enhanced by a matrix-element effect originating in the topological protection of the surface states (Supplementary Information).

\begin{figure*}[t]
\centering
\includegraphics[width=0.7\textwidth]{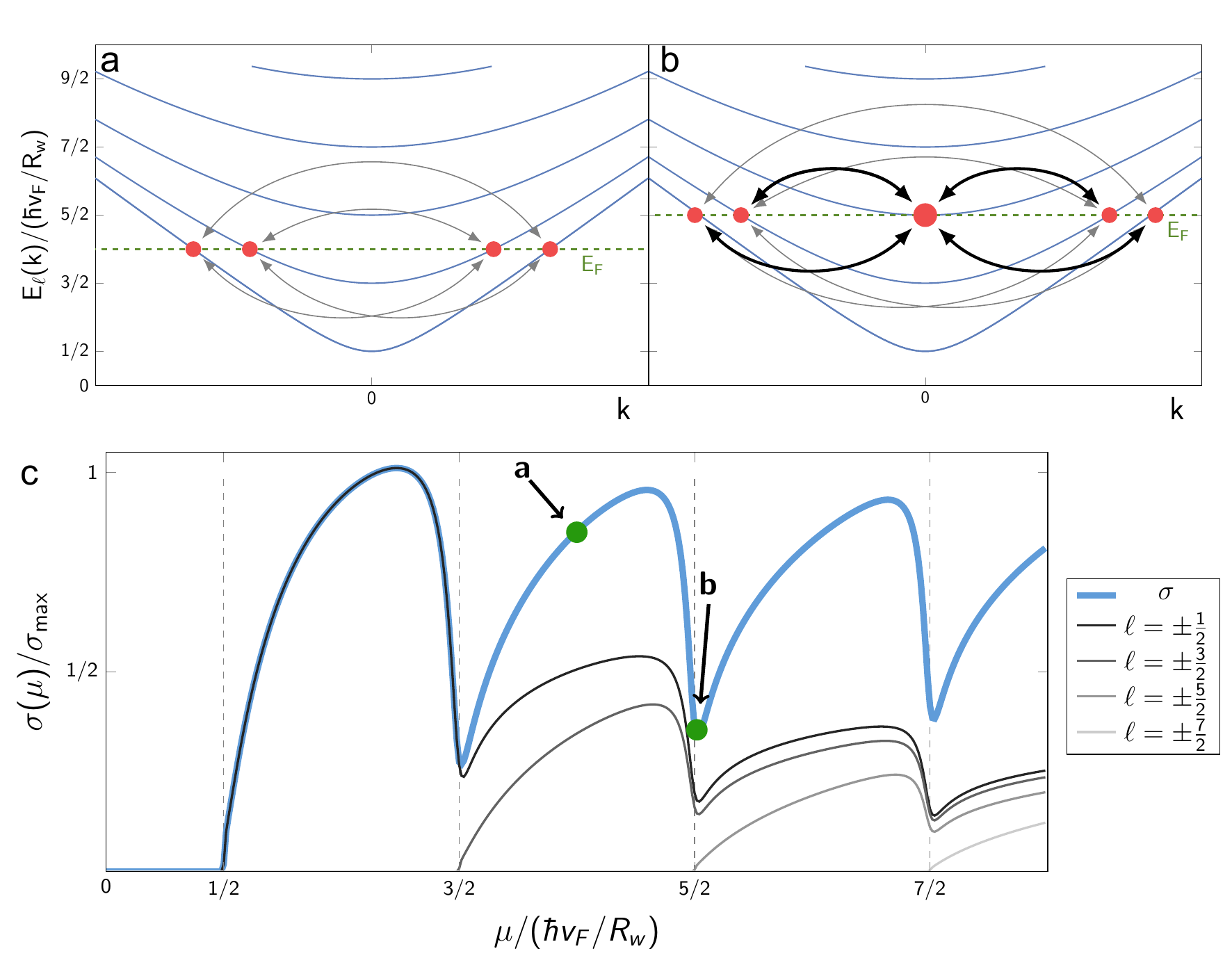}
\caption{\textbf{Scattering processes and conductivity}. \textbf{a, b}, Scattering processes (arrows) between conduction channels for two positions of $\mu$. When $\mu$ is at the bottom of a sub-band, panel \textbf{b}, all other sub-bands scatter at a large rates with the new sub-band due to its diverging density of states, leading to a pronounced minimum in the conductivity.
\textbf{c}, Theoretically-calculated conductivity as function of $\mu$ (parameters as in Fig.~2). Thin black lines display the contribution of each sub-band labeled by $\ell=\pm\frac{1}{2}, \pm\frac{3}{2}, \pm\frac{5}{2}, \pm\frac{7}{2}$, which add up to give the total conductivity (thick blue line). The conductivity of all channels shows pronounced minima at $\mu=\ell \hbar v_F/R_{\rm w}$, when the chemical potential touches the bottom of a new sub-band. 
}
\label{fig:3}
\end{figure*}

Hence, our calculations show that the resistance is expected to show a {\it peak}, each time a sub-band is crossed. This leads to equidistant peaks in Fig.~3, at $\mu=\ell \hbar v_F/R_w$, when the conductivity is plotted as function of the chemical potential $\mu$. In the experiment, however, the gate voltage $V_{\rm G}$, rather than $\mu$, is varied and we observe a super-linear dependence of the spacings of the main peaks (neglecting features of type II). This originates from the fact that the effective capacitance of the nanowire devices (which dictates the $V_{\rm G}$ dependence of the accumulated charge) must be computed from their quantum capacitance $C_{\rm Q}$ and geometric (or galvanic) capacitances $C_{\rm G}$ in series where $C_{\rm Q}$ is proportional to the DOS\cite{Ziegler2018}. In our experiment, $C_{\rm G}$ strongly dominates and the gate voltage directly controls the electron density $n$ ($n \approx C \Delta V_{\rm G}/e$, with $\Delta V_{\rm G}$ measured from the Dirac point), rather than $\mu$. This relation is used for the theory plots in Fig.~2. It also determines the peak positions indicated by dashed lines. We label the position of the peaks by the angular momentum quantum number $\ell$ of the added channel. When the chemical potential reaches the bottom of the first electron or the top of the first hole band ($\ell=\pm \frac{1}{2}$), the charge density is approximately zero in both cases and therefore there is only a single peak in the center for $\ell=\pm \frac{1}{2}$. For large $\ell$, the peak position scales with $\ell^2$, which is peculiar to the sub-bands of Dirac origin.

It is striking that in Figs. 2a-2c the theory can reproduce the essential features of our experiment, in particular the locations of the peaks in the $R(V_{\rm G})$ curves. To visualize the agreement between theory and experiment, we plot in Fig. 2d the rescaled gate voltage values of the peaks, $\Delta V_{\rm G}/V_{0}$,  vs the sub-band index $\ell$, and compare it to the theoretically calculated electron density $n$ at the peak position (in units of $2 \pi/R_{\rm w}$). In these units the rescaling factor is given by 
$V_{0}=\frac{2 \pi e}{R_{\rm w} C}$, where $C$ is the capacitance per length of the wire.
The super-linear behaviour in the $V_{\rm G}$-dependent sub-band crossings and the excellent agreement of theory and experiment is a direct signature of the quantum-confined Dirac surface states, which is observed here for the first time.

It is prudent to mention that the quantum-confined sub-band structure of TI nanowires have been indirectly inferred\cite{Peng2009, Xiu2011, Tian2013, Hamdou2013, Safdar2013, Hong2014, Baessler2015, Cho2015, Arango2016, Jauregui2016, Bhatt2017, Ziegler2018} from the Aharonov-Bohm (AB)-like oscillations of $R$ as a function of the axial magnetic flux $\Phi$, which is due\cite{Zhang2010} to a periodic change in the number of occupied sub-bands at a given $\mu$. In particular, the observation by Cho {\it et al.}\cite{Cho2015} that $R$ at $\Phi$ = 0 takes a maximum when $\mu$ is near the Dirac point and changes to a minimum at some other $\mu$ was consistent with the gapped Dirac cone; however, the $V_{\rm G}$ dependence was not very systematic nor convincing in Ref. \citenum{Cho2015}. A relatively systematic $V_{\rm G}$ dependence of $R$ was recently reported for HgTe nanowires and was carefully analyzed\cite{Ziegler2018}; unfortunately, the Dirac point of HgTe is buried in the bulk valence band, hindering the characteristic super-linear behaviour in the $\Delta V_{\rm G}$ vs $\ell$ relation from observation.

The realization of very thin, bulk-insulating TI nanowires and the observation of the quantum-confined Dirac sub-band structure reported here is crucial for exploring the mesosocpic physics associated with the topological surface states, not to mention their potential for future studies of MZMs. For example, the dependence of the spin degeneracy on the magnetic flux along the nanowires will give us a new tuning knob for mesoscopic transport phenomena, in which the spin-momentum locking can be varied. Also, since one can switch between gapped and gapless 1D Dirac dispersion with a magnetic flux, it would be interesting to address the issue of Klein tunneling. Therefore, the new-generation TI nanowires realized here will open vast opportunities for future studies of topological mesoscopic physics including MZMs.

\section*{Methods}

\subsection{Nanowire synthesis.}
The $\mathrm{(Bi_{1-x}Sb_x)_2Te_3}$ nanowires were synthesized by the VLS method using powders of Bi$_2$Te$_3$ and Sb$_2$Te$_3$ as starting materials in a two-zone 50-mm tube-furnace under a constant Ar flow. The Si/SiO$_2$ substrates were first decorated with suspended 20-nm Au-nanoparticles with the help of Poly-L-Lysine solution and then placed in-between the two zones (set to temperatures $T_1$ and $T_2$) of the furnace. The temperature was first ramped to $T_1$ = 500--510$^{\circ}$C and $T_2$ = 280$^{\circ}$C within 60 minutes, kept at these values for 60 minutes, and finally reduced back to ambient temperature in roughly 4 h, while keeping a constant Ar flow of 600 SCCM.

\subsection{Device fabrication.}
Our gate-tunable four-terminal devices were fabricated on degenerately-doped Si wafers covered by 280-nm thermally-grown SiO$_2$ which acts as a gate dielectric. Gold contact pads and a coordinate system were pre-defined by optical lithography.
The as-grown nanowires were transferred by gently bringing together the surfaces of the prepatterned wafer and the growth substrate, and nanowires suitable for device fabrication were identified by optical microscopy. Per device, five to seven contacts with varying distances were defined by electron beam lithography, which was performed by exposing a PMMA A4 resist layer using a Raith PIONEER Two system. The contact area was cleaned using gentle oxygen plasma treatment and a dip in dilute hydrochloric acid shortly before metallization.
Subsequently, 5-nm-thick Pt was sputter-deposited as a wetting layer and an additional 45-nm-thick Au was deposited by thermal evaporation (device 1 \& 3) or by sputtering (device 2), resulting in the structure schematically shown in Fig.~1e. The contact resistance was well below 1 k$\Omega$ for all of the devices. Following the transport measurements, SEM was used to determine the device geometry and the nanowire diameter.

\subsection{TEM analysis.}
TEM micrographs as well as TEM diffraction patterns were recorded by using a JEM 2200-FS (JEOL) microscope operated at an acceleration voltage of 200 kV. A carbon film supported by a standard copper grid was used as sample carrier for TEM characterization.  Elemental chemical analysis of the samples was done by Energy-Dispersive X-ray Analysis (EDX) performed with a JEOL Dry SD100GV detector.

\subsection{Measurements.}
Transport measurements were performed in a liquid-helium cryostat in the temperature range of 2--300 K. The wafers were glued onto copper sample holders and manually bonded with 50-{$\mu$}m gold wires using vacuum-cured silver paste. For fast measurements, we used a quantum transport measurement system (SPECS Nanonis Tramea) in the low-frequency lock-in mode with the ac current of \SI{100}{nA} at the frequency $f \approx$ 17 Hz, while the device is configured in a conventional four-(device 1\&3) or three-(device 2) terminal geometry. Gate-voltage sweeps were performed at various rates from \SI{0.0125}{V/s} to \SI{0.25}{V/s} while monitoring the sample temperature with a dedicated thermometer using a low-power AC resistance bridge (Lakeshore Model 370).

\subsection{Theoretical calculations.}
We consider the surface states of a quantum wire described by the 2D Dirac equation where antiperiodic boundary conditions in the transverse direction arise from curvature-induced Berry phase effect\cite{Zhang2010}.
Disorder  is modelled by a small density of randomly located local scattering potentials, which is treated within a (non self-consistent) T-matrix approximation, which can be calculated in a fully analytic way, see Supplementary Information. Within our approximation, qualitative features are independent of the density of impurities; however, they do depend on the amplitude of the scattering potential. The Kubo formula was used to calculate the conductivity. Vertex corrections were ignored as a previous study showed that they have only a small, purely quantitative effect\cite{Taskin2017}. Plots of resistivities were obtained from $\rho=1/(\sigma_0+\sigma(\mu))$, where $\sigma_0$ is mainly used to avoid the divergence of $\rho$ when 
$\sigma=0$. It describes the presence of conductance contributions (e.g., from impurity bands on the surface or in the bulk) not taken into account in our approximation. Note that, although the experimental data are shown in resistance $R$,  the theory calculates the resistivity $\rho$, because the transport is assumed to be in the diffusive regime. The dependence of $\mu$ on $\Delta V_{\rm G}$ is computed from $n(\mu)=C_{\rm G}  \Delta V_{\rm G}/e$, where $n(\mu)$ is the electron density along the wire and $C_{\rm G}$ is treated as a fitting parameter. Full details of our calculations are given in the Supplementary Information.

\bibliography{bibliography}

\begin{flushleft} 
{\bf Acknowledgements: }
This project has received funding from the European Research Council (ERC) under the European Union's Horizon 2020 research and innovation programme (grant agreement No 741121) and was also funded by the Deutsche Forschungsgemeinschaft (DFG, German Research Foundation) under CRC 1238 - 277146847 (Subprojects A04, B01, and C02) as well as under Germany's Excellence Strategy - Cluster of Excellence Matter and Light for Quantum Computing (ML4Q) EXC 2004/1 - 390534769.
O.B. acknowledges the support from Quantum Matter and Materials Program at the University of Cologne funded by the German Excellence Initiative.
\end{flushleft} 

\begin{flushleft} 
{\bf Author contributions:}
Y.A. conceived the project. F.M., O.B., D.F. and M.R. performed the growth and device experiments. S.R. performed the TEM analysis. H.F.L, supported by A.R., developed the theory. Y.A., F.M., H.F.L, and A.R. wrote the manuscript with inputs from all authors.
\end{flushleft} 

\begin{flushleft} 
{\bf Competing interests}
The authors declare that they have no competing financial interests.
\end{flushleft} 
 
\begin{flushleft} 
{\bf Correspondence}
Correspondence and requests for materials should be addressed to Y.A.~(ando@ph2.uni-koeln.de)
\end{flushleft}

\clearpage
\onecolumngrid

\renewcommand{\thefigure}{S\arabic{figure}} 

\setcounter{figure}{0}

\begin{center} 
{\Large
\bf Supplementary Information for 
``Quantum confinement of the Dirac surface states in topological-insulator nanowires''
}
\end{center} 
\vspace{2mm}

\maketitle

\baselineskip16pt

\vspace{-20pt}

\section{Experimental description}

\subsection{Growth direction of nanowires}

TEM investigations reveal that the growth direction of the nanowires is parallel to the $\langle 11\bar{2}0 \rangle$ direction, i.e. perpendicular to the c axis of the crystal structure. Since in our TEM characterization the nanowires lie on the supportive carbon carrier, the $\langle 11\bar{2}0 \rangle$ direction is always oriented perpendicular to the optical axis of the microscope. However, the rotation of the nanowires along the growth direction is arbitrary with respect to the support. As a consequence, different orientations perpendicular to the growth direction can be observed by TEM. The c-axis, revealing the distinguished hexagonal symmetry, can only be observed occasionally. This fact also suggests that our nanowires are closer to a cylindrical shape than a ribbon shape.

\subsection{Homogeneity and purity of nanowires}

The chemical composition of the $\mathrm{(Bi_{1-x}Sb_x)_2Te_3}$ nanowire body is homogeneous along the nanowire axis, such that the atomic fraction of each element measured by EDX on six different locations varies only within a few percent (Fig. S1). An incorporation of Au atoms can be excluded within the accuracy of the measurement.

\begin{figure}[h]
\centering
\includegraphics[width=0.6\textwidth]{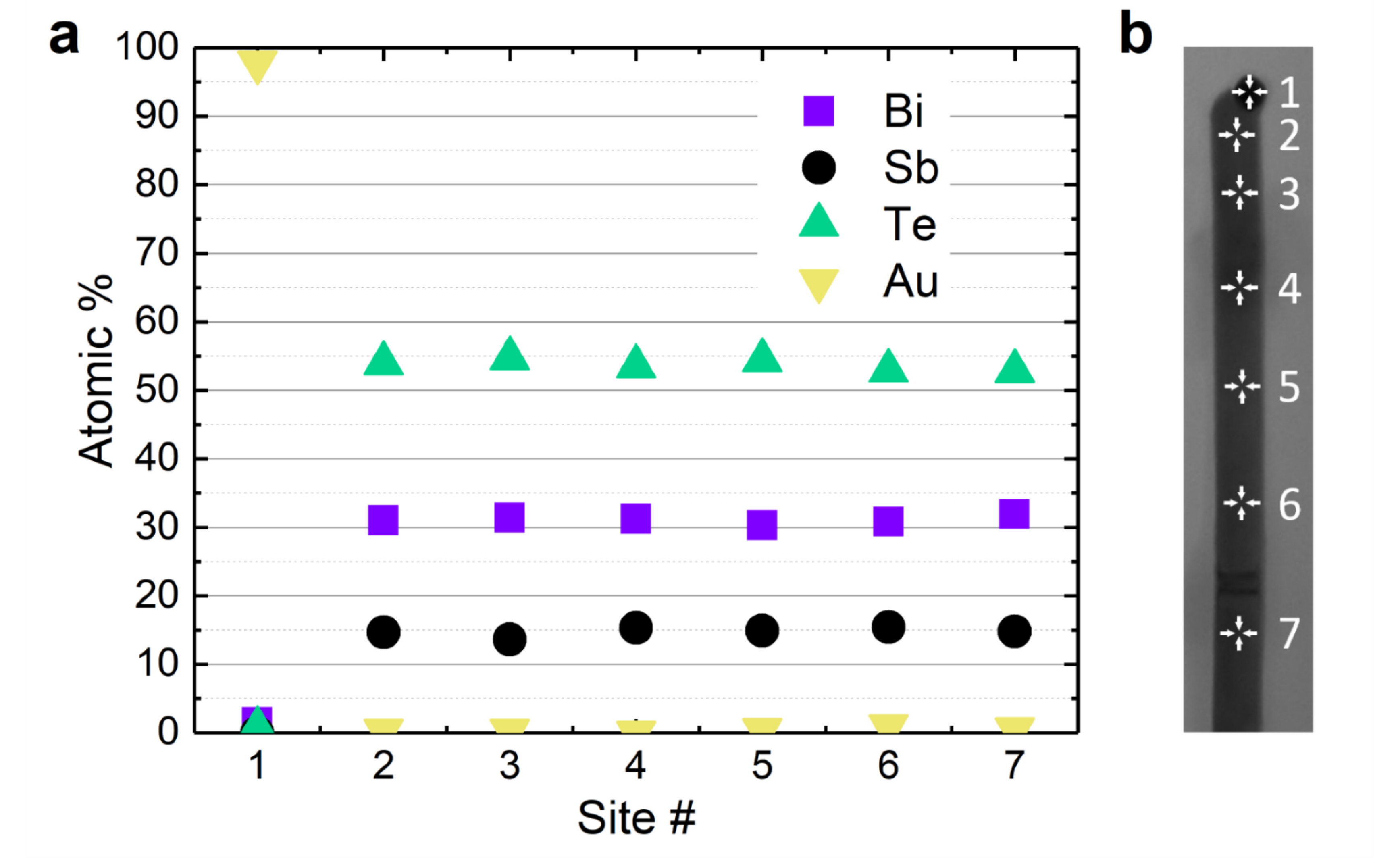}
\caption{\textbf{Series of EDX measurements along the nanowire axis.}
\textbf{a}, Atomic fraction of the elements Bi, Sb, Te, and Au at each location as shown in \textbf{b}.}
\label{fig:EDX}
\end{figure}

\subsection{Identification of the sub-band peaks}

As discussed in the main text, two distinct types of oscillations are seen in the $R(V_{\rm G})$ curves: Large oscillations coming from sub-band crossings (type I) and smaller, time-dependent oscillations (type II). The type-I oscillations are reproducible and most of the peaks are well pronounced. On the other hand, as shown in Fig. S2 for device 3, oscillations of type II are time-dependent with a timescale of a few minutes. Since the gating-curve of device 3 shown in the main text (Fig. 2c) was measured in 80 minutes, the small features are affected by this time dependence.

Actually, sometimes the type-II oscillations make the identification of the type-I peaks complicated, and therefore we have adopted a simple criterion for the identification of the type-I peaks, i.e., the intrinsic peaks coming from the sub-band crossings.

Our criterion is that a pair of peaks should be found at nearly the same $|\Delta V_{\rm G}|$ values for both $+\Delta V_{\rm G}$ and $-\Delta V_{\rm G}$. This particle-hole-symmetry requirement is motivated by the observation from the angle-resolved photoemission spectroscopy on $\mathrm{(Bi_{1-x}Sb_x)_2Te_3}$ that the Dirac cone in this system is essentially particle-hole symmetric \cite{Zhang2011}. This simple criterion greatly helps us to reject spurious peaks and identify intrinsic peaks of the type-I origin, which turn out to agree well with the theoretical calculations as shown in the main text.

\begin{figure}[t]
\centering
\includegraphics[width=0.6\textwidth]{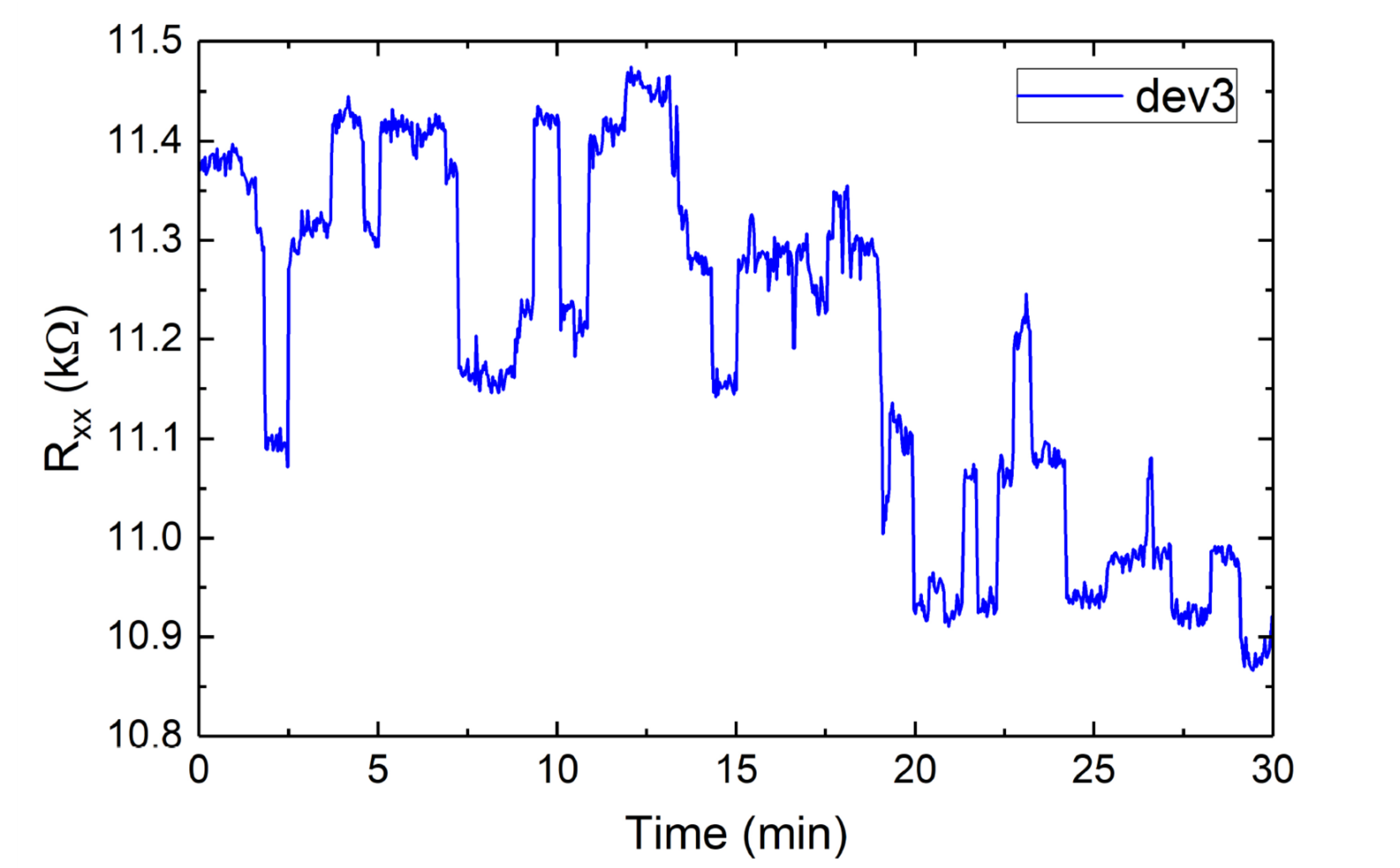}
\caption{Time-dependent resistance fluctuations of type II measured on device 3.}
\label{fig:t-fluc}
\end{figure}

\subsection{Geometric and quantum capacitances}

In general, the gate voltage required to provide the nanowire with additional charge $Q$ can be expressed as
\begin{equation} \label{eqn:capacitance}
V_\mathrm{G}=\frac{Q}{C_\mathrm{G}} + \frac{Q}{C_\mathrm{Q}},
\end{equation}
where $C_\mathrm{G}$ is the geometric capacitance of the device and $C_\mathrm{Q}$ is the quantum capacitance,
or, equivalently, as $e V_\mathrm{G}=\frac{e^2}{C_\mathrm{G}} n(\mu)-\mu$, where $e<0$ is the electron charge, $\mu$ the chemical potential, and $n(\mu)$ the electron density. In our experiments, a change of the gate voltage by $30$\,V leads to changes of the chemical potential by less then $100$\,meV. Therefore, the effect of the geometric capacitance dominates by more than two orders of magnitude and we can safely approximate
$V_\mathrm{G} \approx \frac{1}{C_\mathrm{G}}  e n(\mu)$.

The geometric capacitance $C_\mathrm{G}$ was found to be 2 -- 6 pF/m in our experiments. Note that $C_\mathrm{G}$ depends on materials properties and the device geometry, and for the present gate geometry, a value of $C_\mathrm{G} \approx$ 50 pF/m would be expected from numerical simulations \cite{Wunnicke2006}. However, our nanowires are not metallic as is typically assumed in simulations, and hence the result of such numerical simulations should be taken as an upper bound for $C_\mathrm{G}$. In addition, the effective capacitance is further affected by  parasitic capacitances due to the leads, as well as by the finite length of the nanowire. For these reasons we  treat the value of the capacitance as a phenomenological parameter.

\section{Theoretical description}
\subsection{Hamiltonian of a disordered TI nanowire}
The Hamiltonian for the surface of a cylindrical TI nanowire of radius $R_w$ depends only on momentum around the wire, $\hat{p}_\phi=\hat{L}_\phi/R_w=-i v_F \partial_{\phi}/R_{\rm w}$, and momentum along the wire, $\hat{p}_x=-i\hbar \partial_x$, and is given by \cite{Zhang2010} 
\begin{equation}
H_{\rm surf}=i\hbar v_F \left(\frac{\sigma_x}{R_{\rm w}} \partial_{\phi}-(\sin(\phi) \sigma_z-\cos(\phi) \sigma_y)\partial_x\right).
\end{equation}
This can be simplified by applying a spinor-rotation by $\phi$ about the x-axis, {\it i.e.}, the unitary transformation $U_x(\phi)=\exp(-i \phi \sigma_x/2)$. The result is the Hamiltonian
\begin{equation}
H_0=i\hbar v_F \left(\frac{\sigma_x}{R_{\rm w}} \partial_{\phi}-\sigma_y\partial_x\right).\label{H0}
\end{equation}
In the presence of a flux $\Phi$ threaded along the wire, $\partial_\phi$ is replace by $\partial_\phi-i \eta$ with $\eta=\Phi/\Phi_0$.
Importantly, since spinor rotations are $4\pi$ periodic, the eigenfunctions of this Hamiltonian must obey anti-periodic boundary conditions, resulting from $U_x(\phi)=-U_x(\phi+2\pi)$. As such the wave function can be written in the general form $\psi_{\ell}(k)=e^{i(k x+\ell \phi)}\boldsymbol{\xi}$,
where the anti-periodic boundary conditions require the total angular momentum to satisfy $\ell=\pm 1/2, \pm 3/2, \dots$ and $\boldsymbol{\xi}$ is a spinor encoding the spin-structure of the surface state within this rotated system.

Due to the finite geometry along the circumference of the wire the eigenenergies of the Hamiltonian are quantised 
\begin{equation}
E_{\ell,\pm}(k) =\pm\hbar v_F\sqrt{k^2+\left(\frac{\ell-\eta}{R_{\rm w}}\right)^2}.
\end{equation}
The band edge of each band is for $\eta=0$ given by $\varepsilon_{\ell,\pm}=\pm \hbar v_F \ell/R_{\rm w}$. We define $\varepsilon_\ell=\varepsilon_{\ell,+}$.

Disorder is modelled by impurity potentials located at random positions $\vec r_i$, 
$V_i(\vec r)=u_0\mathbb{1}\delta(\vec r-\vec r_i)$, where $\vec r$ and $\vec r_i$ are 2d coordinates on the surface of the wire. 


\subsection{Density of states and charge density}
The free retarded Green's function -- which is a matrix -- has the form
\begin{equation}
[\vec{G}^0]_{\alpha\beta}(\mu, k,\ell)=(\mu \mathbb{1}-H_0(k,\ell)+i \delta\mathbb{1})^{-1}_{\alpha\beta},
\end{equation}
where we take the matrix inverse. Here $\delta$ acts as some overall background scattering rate due degrees of freedom (e.g. impurity bands) not described by our Hamiltonian. It has the effect of broadening the 
divergence of the density of states at the bottom of each sub-band.
 
The local Green's function of each $\ell$ resolved sub-band is given by
\begin{equation} 
\vec \Delta(\mu,\ell)=\frac{1}{2\pi R_{\rm w} v_F}\int \frac{dk}{2\pi \hbar} \vec{G}^0(\mu,k,\ell) =-\frac{(\mu+i \delta) \mathbb{1}-\varepsilon_\ell \sigma_x}{4\pi v_F R_{\rm w} \hbar \sqrt{\varepsilon_\ell^2-(i\delta+ \mu)^2}}.\label{delta}
\end{equation}

From the Green's function we can obtain the local density of states (at a fixed position at the surface of the wire)
\begin{equation}
\rho(\mu)=-\frac{1}{\pi}\sum_{\ell=\pm\frac{1}{2},\pm\frac{3}{2},\dots} {\rm Im}\left\{{\rm Tr}\; \vec \Delta(\mu,\ell) \right\} = \frac{1}{2\pi^2 R_{\rm w} v_F \hbar} \sum_{\ell=\pm\frac{1}{2},\pm\frac{3}{2},\dots} {\rm Im}\{\frac{\mu+i \delta}{\sqrt{\varepsilon_\ell^2-(i\delta+ \mu)^2}}\},
\end{equation}
which, as should be expected, tracks the 2d Dirac density of states $\rho^{\rm 2d}(\mu)=|\mu|/2\pi v_F^2 \hbar^2$.

From the density of states we can also obtain the charge density along the wire. It is given by
\begin{equation}
n(\mu)=2\pi R_{\rm w}\int^\mu_{0} d\mu' \rho(\mu')=-\frac{1}{\pi v_F \hbar}{\rm Im}\left\{\sum_{\ell=\pm\frac{1}{2},\pm\frac{3}{2},\dots} \sqrt{\varepsilon_\ell^2-(\mu+i \delta)^2}\right\}.
\end{equation}
Here $n(\mu)$ is zero at the Dirac point and counts the total charge density per length of wire. For $\mu>0$ this means counting the contribution of all bands above the Dirac point up to the chemical $\mu$. Correspondingly, for $\mu<0$ counting the charge of all holes contributed by bands below the Dirac point, in other words $n(\mu)$ is negative for $\mu<0$. It is shown as a grey line in Fig.~2d of the main text.

\subsection{T-matrix approximation}
The full disorder averaged Green's function is 
\begin{equation}
[\vec{G}]_{\alpha\beta}(\mu, k,\ell)=(\mu \mathds{1}-H_0(k,\ell)+\vec \Sigma(\mu))^{-1}_{\alpha\beta}. \label{green},
\end{equation}
where $\vec \Sigma(\mu)$ is the full self-energy from all scattering processes, including disorder and the other scattering processes modelled by the factor $i \delta$ (see above).

To approximate the self-energy contribution due to disorder scattering we use the full T-matrix calculated at first order in impurity density $n_{\rm imp}$  (i.e. the non-crossing approximation). Within this approximation the self-energy, which is also a matrix, is \cite{Taskin2017}
\begin{align}
[\vec{\Sigma}]_{\alpha\beta}(\omega)=n_{\rm imp} u_0 \left\langle \left( \mathbb{1}-\frac{u_0}{2\pi R_{\rm w}} \sum_{\ell=\pm\frac{1}{2},\pm\frac{3}{2},\dots}\int \frac{dk}{2\pi \hbar v_F} \vec{G}^0(k,\mu,\ell)\right)^{-1} \right\rangle_{\rm imp}\\
=n_{\rm imp} u_0 \left( \mathbb{1}-u_0\sum_{\ell=\pm\frac{1}{2},\pm\frac{3}{2},\dots}\Delta\left(\mu,\ell\right)\right)^{-1}.\nonumber
\end{align}
For small $\delta$ the $\vec \Delta(\mu,\ell)$, from \eqref{delta}, only has a substantial imaginary part for $|\mu|>|\varepsilon_\ell|$, this means that a conductivity channel only contributes when the corresponding sub-band has an occupation of electrons (holes) for the upper (lower) Dirac cone. Further the off-diagonal components of $\vec \Sigma$ are exactly zero for diagonal impurities since the off-diagonal contributions of positive and negative $\ell$ cancel.

The self-energy can also  be calculated self-consistently by replacing $\mu \rightarrow \mu+\Sigma_{11}$ and $\varepsilon_\ell \rightarrow \varepsilon_\ell+\Sigma_{12}$, where $\Sigma_{11}$ and $\Sigma_{12}$ are the diagonal and off-diagonal components, respectively. Self-consistency has a similar impact on the self-energy as $\delta$, that is, the breadth of the peaks in scattering rate become broadened in proportion to the self-energy itself. Ultimately this means that peaks at high sub-band index become ``washed out'' when the self-energy is of the same order of magnitude as the sub-band spacing. This is not of relevance for the experimental situation discussed here where the mean free path is several times the wire radius, but would limit the number of peaks seen in thicker or dirtier wires where the ratio of mean free path and radius can be substantially smaller.


\subsection{Conductivity}
The conductivity is found using the Kubo formula to calculate the current-current correlation function in linear response. The current operator along the wire is given by $j_x=e\partial H/\partial k=-i \hbar v_F e \sigma_y$. We neglect vertex corrections which previous studies have shown only have a quantitative effect on overall conductivity\cite{Taskin2017}. Within this scheme the DC conductivity is given by
\begin{align}
\sigma&=\frac{e^2 v_F^2 \hbar^2}{\pi} \frac{1}{2\pi R_{\rm w}} \sum_{\ell=\pm\frac{1}{2},\pm\frac{3}{2},\dots} \int d\omega\;n_F'(\omega)\int \frac{dk}{2\pi \hbar}{\rm Tr}\left( \sigma_y \vec G(\omega, k,\ell) \sigma_y \vec G^\dagger(\omega, k,\ell)\right)= \sum_{\ell=\pm\frac{1}{2},\pm\frac{3}{2},\dots} \sigma_\ell(\mu),
\end{align}  
where in general the contribution $\sigma_\ell$ is given by
\begin{equation}
\sigma_\ell(\mu)=\frac{e^2 v_F }{(2\pi)^2 R_{\rm w}}\frac{1}{\tilde{\mu}\Gamma_{11} -\tilde{\varepsilon_\ell} \Gamma_{12} }{\rm Im}\left\{\frac{\lambda}{\sqrt{\xi}}-\frac{\lambda^\dagger}{\sqrt{\xi^\dagger}} \right\} 
\approx \theta(\mu^2-\varepsilon_l^2)\frac{e^2 v_F}{2\pi^2 R_{\rm w} |\Gamma_{11}|} \frac{\mu^2-\varepsilon_l^2}{\mu \sqrt{\mu^2-\varepsilon_l^2}},\label{cond}
\end{equation}
where $\vec \Gamma={\rm Im}\vec \Sigma$ is the (spin-resolved) scattering rate, $\tilde{\mu}=\mu+{\rm Re}{\Sigma_{11}}$, $\tilde{\varepsilon}_\ell=\varepsilon_\ell+{\rm Re}{\Sigma_{12}}$, and the factors $\xi=-(\mu+i \Gamma_{11})^2+(\varepsilon_\ell+i\Gamma_{12})^2$ and $\lambda=-\tilde{\varepsilon}_\ell^2-i \tilde{\varepsilon}_\ell \Gamma_{12}+i \tilde{\mu}  \Gamma_{11}+\tilde{\mu}^2$ ensure the channel only contributes to conductivity when $|\mu| \gtrsim |\varepsilon_\ell|$. The approximation in \eqref{cond} is valid for small self-energies and diagonal impurities. 

\begin{figure}[b]
\centering
\includegraphics[width=0.8\textwidth]{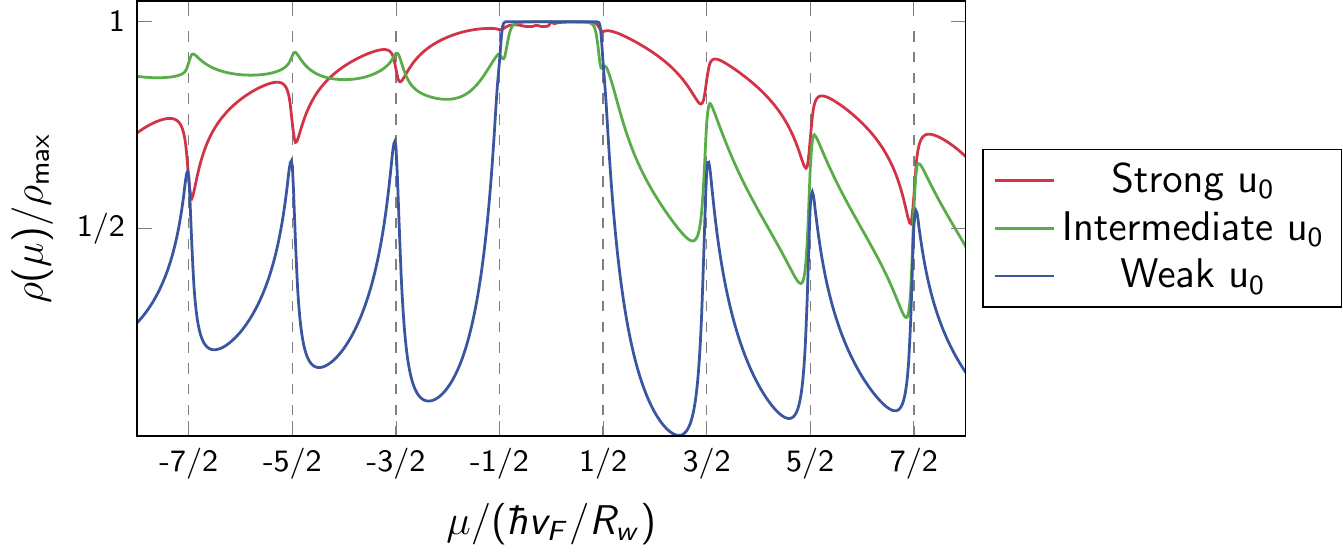}
\caption{\textbf{Resistivity as a function of chemical potential in the weak, intermediate, and strong scattering regimes}: A weak scatterer, $\frac{u_0}{\hbar v_F R_{\rm w}}=0.1$, $n_{imp}=20/R_{\rm w}^2$, as in the main text, is shown as the blue curve. The green curve, $\frac{u_0}{\hbar v_F R_{\rm w}}=1$,  $n_{imp}=2/R_{\rm w}^2$, describes scatterers of intermediate strength,  the red curve, a strong scatterer $\frac{u_0}{\hbar v_F R_{\rm w}}=20$, $n_{imp}=0.1/R_{\rm w}^2$. Strong scattering is capable of turning resitivity peaks at sub-band edges into resitivity dips, however such a situation is unlikely (see text).}
\label{fig:res}
\end{figure}
\subsection{Parameters for theory plots}
An important parameter for our theory is $u_0$, the strength of the impurity potential. In Fig.~\ref{fig:res} we show the resistivity as function of gate voltage for weak, intermediate and strong impurity potentials.
Only for weak impurity potentials, $\frac{u_0}{\hbar v_F R_{\rm w}}\ll 1$, peaks occur at the band edges and particle- and hole doping are almost equivalent. For intermediate impurity strength, $\frac{u_0}{\hbar v_F R_{\rm w}}\sim 1$, the curve is highly asymmetric. For $\frac{u_0}{\hbar v_F R_{\rm w}}\gg 1$, the particle-hole symmetry is restored but instead of peaks one obtains dips.

We find that only weak $u_0$ fit the experimental data. This is expected because the wavefunction of the conduction channel wraps around the circumference of the wire, effectivly diluting the impurity by a factor $1/(2 \pi R_{\rm w})$. To be precise: Weak scattering can be defined by $u_0 \bar \rho \ll 1$, where $\bar \rho \sim \frac{1}{2\pi v_F \hbar R_{\rm w}}$ is the typical density of states in the system for small $\ell$. Assuming that the width of the scattering potential is of the order of the lattice constant $a$, the weak scattering limit is reached when the amplitude, $V_0=u_0/a^2$, of the scattering potential
fulfils $V_0 \ll \frac{v_F}{a \hbar} \frac{2 \pi R_{\rm w}}{a}$. Since in our experiments the circumference is many lattice constants, $\frac{R_{\rm w}}{a} \sim 100$, the weak scattering limit is generically realised.

Therefore, for the plots in Fig.~2 and Fig.~3 of the main text we use $u_0=0.1 v_F \hbar R_{\rm w}$, $n_{\rm imp}=20/R_{\rm w}^2$ and $\delta=0.025 v_F \hbar/R_{\rm w}$, see also method section. 
This reproduces the order of magnitude $\sim 5$ nm/k$\rm \Omega$  of the fluctuations in wire conductivity found experimentally.

While the main contribution to the peaks at the subband edges arised from the diverging density of states, this effect is further enhance by a matrix-element effect closely related to the topological protection of TI surface states. Here the spin orientation is locked to the propagation direction and scattering from $\vec k$ to $-\vec k$ is prohibited by time-reversal symmetry. In our nanowires, $k_y=\ell/R_{\rm{w}}$ is quantized, and the 1D bands are doubly degenerate as for each $k_x=k$, two transverse momenta, $\pm k_y$, are possible. For $k_x \gg k_y$, however, the 2D vectors $\vec k=(k_x,\pm k_y)$ and $\vec k=(-k_x,\pm k_y)$ are almost antiparallel, leading to a suppression of the scattering rate by the factor $(k_y/k_x)^2$. This matrix element effect (encoded in the $2 \times 2$ matrix structure of the T-matrix) substantially suppresses  backscattering among occupied channels with $\ell/R_{\rm{w}}\ll k_x$ 
relative to the scattering to a newly opened channel with a large $k_y$.

\subsection{Electron-hole puddles}
Several physical effects have not been taken into account by our theoretical treatment. For example, we do not reproduce the type II fluctuations which likely arise from extra charge traps in the experimental setup. We also assume scattering from local impurities, but there will be also long-range potentials arising from randomly distributed positive and negative charges in our compensation doped samples of (Bi$_{1-x}$Sb$_x$)$_2$Te$_3$. In bulk systems, these lead to the formation of electron-hole puddles \cite{Borgwardt2016,Breunig2017} both in the bulk and in the surface states \cite{Knispel2017} of TIs. Estimates from Ref.~\cite{Boemerich2017} suggest that bulk puddles are absent in TI nanowires with a radius of less than $\sim 100$\,nm. But also in this case, one can expect substantial fluctuations of the chemical potential on the surface of the TI. For surface states of bulk samples of BiSbTeSe$_2$, fluctuations of the local chemical potential with an amplitude of  $~20\,$meV on a length scale of $50$\,nm have been measured by scanning tunneling spectroscopy \cite{Knispel2017}. While the size of these fluctuations will be reduced for nanowires, surface puddles can be expected to be of quantitative importance (perhaps, also for type II fluctuations).

\newpage

\end{document}